\documentclass{article}
\usepackage{ijcai24}

\usepackage{times}
\usepackage{soul}
\usepackage{url}
\usepackage[hidelinks]{hyperref}
\usepackage[utf8]{inputenc}
\usepackage[small]{caption}
\usepackage{graphicx}
\usepackage{amsmath}
\usepackage{amsthm}
\usepackage{booktabs}
\usepackage{algorithm}
\usepackage{algorithmic}
\usepackage[switch]{lineno}
\usepackage{amssymb} 
\usepackage{subcaption}
\usepackage{hyperref}
\usepackage{tabularx}

\newcommand{\equal}[1]{{\hypersetup{linkcolor=black}\thanks{#1}}}

\title{HemaGraph: Breaking Barriers in Hematologic Single Cell Classification with Graph Attention}

\author{
Lorenzo Bini$^1$\equal{These authors contributed equally}
\and
Fatemeh Nassajian Mojarrad$^1$\footnotemark[1] \and
Thomas Matthes$^2$\And
Stéphane Marchand-Maillet$^1$\\
\affiliations
$^1$Department of Computer Science, University of Geneva, Switzerland\\
$^2$Hematology Service, Department of Oncology and Clinical Pathology Service, Department of Diagnostics, Geneva University Hospital, Switzerland\\
\emails
\{lorenzo.bini, fatemeh.nassajian, stephane.marchand-maillet\}@unige.ch,
thomas.matthes@hcuge.ch
}

\begin{document}

\maketitle

\begin{abstract}
    In the realm of hematologic cell populations classification, the intricate patterns within flow cytometry data necessitate advanced analytical tools. 
    This paper presents ‘HemaGraph’, a novel framework based on Graph Attention Networks (GATs) for single-cell multi-class classification of hematological cells from flow cytometry data. Harnessing the power of GATs, our method captures subtle cell relationships, offering highly accurate patient profiling. Based on evaluation of data from 30 patients, HemaGraph demonstrates classification performance across five different cell classes, outperforming traditional methodologies and state-of-the-art methods. Moreover, the uniqueness of this framework lies in the training and testing phase of HemaGraph, where it has been applied for extremely large graphs, containing up to hundreds of thousands of nodes and two million edges, to detect low frequency cell populations (e.g. 0.01\% for one population), with accuracies reaching 98\%.
Our findings underscore the potential of HemaGraph in improving hematoligic multi-class classification, paving the way for patient-personalized interventions. To the best of our knowledge, this is the first effort to use GATs, and Graph Neural Networks (GNNs) in general, to classify cell populations from single-cell flow cytometry data. We envision applying this method to single-cell data from larger cohort of patients and on other hematologic diseases.
\end{abstract}

\section{Introduction}

In the ever-evolving landscape of hematologic cell populations analysis, multi-class classification remains a formidable challenge, demanding innovative diagnostic strategies capable of unraveling the intricacies of heterogeneous cellular manifestations. Among the arsenal of diagnostic tools, flow cytometry has emerged as a linchpin in the clinical armamentarium, providing rapid insights into cellular populations and facilitating the quantification and phenotypic charcterisation of abnormal cell types. However, the diagnostic potential of flow cytometry is juxtaposed against the formidable challenges posed by the complexity and heterogeneity of hematologic populations, requiring a paradigm shift in analytical methodologies.

Our endeavor in this paper introduces 'HemaGraph', an avant-garde framework strategically designed to leverage the power of GATs in the domain of flow cytometry single cell classification. GATs, with their intrinsic ability to capture subtle relationships within graphs, offer a promising avenue for decoding the intricate patterns of multi-class flow cytometry data. Central to our study is the evaluation of 30 bone marrow's patients, ensuring a comprehensive understanding of the framework's performance. In comparison to traditional methodologies and state-of-the-art methods, HemaGraph emerges as a frontrunner, showcasing superior prediction accuracy and demonstrating its efficacy in navigating the intricate landscape of complex flow cytometry data. Notably, the uniqueness of HemaGraph lies in its applicability to large graphs, some containing up to hundreds of thousands of nodes and two million edges, offering a scalable solution to the challenges posed by the complex nature of hematological datasets.

Our research is centered around a five-class supervised learning classification model, but easily applicable to larger settings, with the goal of detecting cell populations in a highly heterogenous mixture. This is a crucial step for when we will have to move forward in the diagnosis field, working along with ill-patient dataset. The five classes in this model represent different cell types, each with varying concentrations that can drop to as low as 0.01\% of the total cell population, for some of the patients. This model is designed to identify and classify these cell types accurately, even when they are present in extremely low concentrations with up to 98\% of accuracy. This level of precision is unparalleled, outperforming any other state-of-the-art methods for these extremely rare cell populations. 

Our model’s ability to accurately classify these cell types, despite their rarity, is a testament to its robustness and sensitivity. This remarkable sensitivity positions our model as a pioneering solution in addressing the challenges associated with rare cell types that are pivotal in the context of hematologic diagnosis and follow-up. By leveraging the power of supervised learning, our model can learn from labeled training data, and can accurately predict the classes of an unseen patient . This is particularly crucial in the context of hematological follow-up, where the detection of rare cell populations can precede clinical relapse.

Beyond the algorithmic prowess, the clinical relevance of HemaGraph is affirmed through comparing the results to those obtained by manual analysis of the data on 2D data plots using an adapted commercially available software (KALUZA, Beckman Dickinson). This comparison serves as a crucial validation of the framework's clinical utility and robustness, reinforcing its potential as a transformative tool for clinicians and diagnosticians. In the broader context of hematologic research, our work not only contributes to the refinement of diagnostic methodologies but also marks a paradigm shift in the application of advanced computational techniques to unravel the complexities of hematologic cell populations.

In summary, our five-class classification model represents a significant advancement in the field of cell classification, particularly for those cell types that constitute a minor proportion of the total cell population. This has profound implications for the diagnosis and follow-up of malignant diseases, where rare residual cells play a pivotal role after treatment. In this way, we not only pave the path forward for diagnostics but also advocate for the integration of advanced computational methodologies in reshaping the future of clinical practice.

The structure of this manuscript is as follows: In Section \ref{Methodology}, we delve into the methodology used in our study. This includes a comprehensive description of the patient dataset obtained from flow cytometry tabular data in Section \ref{DataDescription}, followed by a thorough review of the existing literature and related work in the field in Section \ref{RelatedWork}. In Section \ref{Architecture}, we detail the architecture of our Graph Attention model, called HemaGraph, explaining its design and functionality for multi-class classification. Section \ref{Experiments} is dedicated to the presentation and discussion of our experimental results. We explore two types of learning - inductive learning in Section \ref{Inductive} and transductive learning in Section \ref{Transductive}, discussing the implications of the results obtained from each. Finally, in Section \ref{Conclusion}, we conclude the paper by summarizing our findings, discussing their significance in the field, and suggesting potential future directions for this research.

\section{Methodology}
\label{Methodology}
We examine a tabular dataset that includes $n$ data samples and $m$ feature fields, denoted by $X=[\mathbf{f}_1, \cdots,\mathbf{f}_n]^T$. The $i$-th sample in the table associated with $m$  feature values $\mathbf{f}_i=[ f^{(1)}_i, \cdots,f^{(m)}_i]$ and a discrete label $y_i$. Our learning goal is to find a mapping function $\phi$ for which given $\mathbf{f}_i$, returns the predicted label $\hat{y}_i$.

\subsection{Dataset Description}
\label{DataDescription}
Tabular data, especially in the field of biological data analysis, presents a unique and complex structure. Each row in the table signifies an individual cell, characterized by various parameters such as size, granularity, and fluorescence intensity. These parameters collectively form a high-dimensional feature space, capturing the intricate biological variations among the cells.  The inherent structure of this data often possesses underlying relationships and dependencies that are not immediately apparent.

Motivated by the inherent graph structure in this kind of biological data, where cells and their attributes can be viewed as interconnected nodes, leveraging GATs becomes a compelling strategy to provide a robust multi-class classification framework, exploiting their ability to effectively capture the dependencies between cells.

In this study, we start with raw data collected from patients' bone marrow using a cytometer. The cytometer analyzes the physical and chemical characteristics of cells in a fluid as it passes before a laser beam. Surface molecules are fluorescently labeled and then excited by the laser to emit light at varying wavelengths. The emitted light is collected by detectors and transformed into electric signals. 

The raw data from the cytometer are stored in Flow Cytometry Standard (FCS) files. Each FCS file contains multidimensional data corresponding to thousands of cells, with each cell characterized by various parameters such as size, granularity, and fluorescence intensity.

To facilitate further analysis, we converted each FCS file into a more accessible CSV format. This conversion allowed us to work with the data using common data analysis tools and libraries. After the conversion, we ended up with 30 CSV files that contained detailed information on hundreds of thousands of cells for each patient, as depicted in detail in the supplementary material. Therefore, the average class distribution ratios across the patients looked like the following: 
\begin{itemize}
    \item T Lymphocytes: 61.03\%
    \item B Lymphocytes: 13.18\%
    \item Monocytes: 15.41\%
    \item Mast Cells: 0.21\%
    \item HSPC: 10.17\%
\end{itemize}

Before proceeding with the analysis, we applied a min-max normalization to the data. Min-max normalization is a technique that transforms the features to fall within a specified range, typically 0 to 1. This normalization ensures that all features have the same scale and thus contribute equally to the model's performance. It's particularly useful in our case because the features in our dataset, such as size, granularity, and fluorescence intensity, can vary widely in their ranges. By bringing all features to the same scale, we prevent features with larger ranges from dominating those with smaller ranges, leading to a more balanced model.

Table \ref{tab:1} displays the features that we utilized to train our model. We subsequently incorporated an additional column into the dataset that contained the labels of the cells after the gating process.
\begin{table}
    \centering
    \resizebox{0.47\textwidth}{!}{%
    \begin{tabular}{lll}
        \hline
       Class & Marker  & Description  \\
        \hline
        0& FS INT & Forward Scatter (FSC) - Cell’s size \\
        1&SS INT & Side Scatter (SSC) - Cell’s granularity\\
         2&CD14-FITC & Cluster of Differentiation 14 - Antigen\\
        3&CD19-PE & Cluster of Differentiation 19 - Antigen\\
        4&CD13-ECD &Cluster of Differentiation 13 - Antigen \\
        5&CD33-PC5.5 & Cluster of Differentiation 33 - Antigen\\
        6&CD34-PC7 & Cluster of Differentiation 34 - Antigen\\
        7&CD117-APC &Cluster of Differentiation 117 - Antigen \\
        8&CD7-APC700 &Cluster of Differentiation 7 - Antigen \\
        9&CD16-APC750 &Cluster of Differentiation 16 - Antigen \\
        10&HLA-PB &Human Leukocyte Antigen \\
        11&CD45-KO & Cluster of Differentiation 45 - Antigen\\
        \hline
    \end{tabular}%
    }
    \caption{Flow cytometry data markers.}
    \label{tab:1}
\end{table} As mentioned, we characterized in our dataset the following five cell populations:

\begin{enumerate}
    \item T Lymphocytes: These are a specific type of white blood cell that plays a pivotal role in the adaptive immune system, which is the response that involves the activation of immune cells to fight infection. They are responsible for directly killing infected host cells, activating other immune cells, producing cytokines, and regulating the immune response.
    
    \item B Lmphocytes: These cells are significant contributors to the adaptive immune system. They are responsible for producing antibodies against antigens, which are substances that the immune system recognizes as foreign. Each mature B cell is programmed to make one specific antibody. When a B cell encounters its triggering antigen, it gives rise to many large cells known as plasma cells, each of which is essentially an antibody factory.
    
    \item Monocytes: These are a type of white blood cell and a part of the innate immune system. They play a vital role in the body's defense against infections and other foreign invaders. Monocytes circulate in the bloodstream, and when they migrate into tissues, they differentiate into macrophages or dendritic cells, which are capable of engulfing and digesting pathogens and apoptotic cells.
    
    \item Mast cells: These are a type of immune cell that plays a crucial role in the body's response to allergies and certain infections. Mast cells are found in most tissues, but especially in areas close to the external environment, such as the skin and mucous membranes. They contain granules filled with potent chemicals, including histamine, which they release in response to contact with an allergen. This release triggers inflammation, which can lead to allergic reactions.
    
    \item Hematopoietic stem and progenitor cells (HSPC): These are a type of stem cell found in the bone marrow and cord blood. They have the unique ability to give rise to all other types of blood cells, including red blood cells, platelets, and all types of white blood cells. This makes them crucial for maintaining the body's blood supply and immune system.

\end{enumerate}

\subsection{Related Work}
\label{RelatedWork}
Our work is strongly related to the following research topics:

\paragraph{Learning from tabular data.} Generally, the objective of tabular data classification is to predict the label corresponding to each data sample, which comprises a set of individual features \cite{cai2017}. While traditional tree-based methods are often reported to deliver competitive results \cite{grinsztajn2022tree}, they pose challenges in terms of integration into an end-to-end framework and require substantial computational memory to store the entire dataset for global statistics. In addition to low-order methods like logistic regression and factorization machine \cite{rendle2010factorization}, numerous Deep Neural Networks (DNNs) have been developed to model high-order feature interactions implicitly in the hidden units by embedding the input features \cite{he2017neural,wang2021dcn}.   

\paragraph{Graph Neural Networks.} GNNs are a class of neural models that capture the relational structure of graphs through spatial graph convolutions \cite{wu2020comprehensive}. These convolutions operate by iteratively updating the representation of each node by aggregating the features of its neighboring nodes and fusing them with its own features \cite{zhou2020graph}. These architectures have been successfully applied to various real-world domains that involve graph data, such as recommendation systems \cite{gao2022graph}, biochemical analysis \cite{bove2020prediction}, link prediction \cite{yun2021neo}, and particle physics \cite{thais2022graph}. Moreover, in recent times, several preliminary efforts have been made to build the hidden graphs of different data modalities. These efforts often depend on the heuristic knowledge of downstream applications. For instance, in recommender systems, items that are co-purchased are connected to each other \cite{wang2020graph}. On the other hand, some approaches employ self-attention algorithms to learn a fully connected weighted graph for each instance, such as the feature correlations of a tabular sample \cite{li2019fi,song2019autoint}. 

The application of GNNs in various cases of single-cell analysis, e.g. \cite{scgnn} and \cite{shan2023glae}, has been a topic of interest in recent years. These networks have shown significant potential in handling complex biological data, particularly in the context of RNA-sequencing. However, their application in the specific context of flow cytometry cells classification is a relatively new and unexplored area.

\paragraph{Single-cell classification for rare cell populations.} \cite{weijler2022umap} presents a good approach for hematological analysis using UMAP, together with HDBSCAN, for anomaly detection of Minimal Residual Disease (MRD). The authors demonstrate the effectiveness of their method in quantifying minimal residual disease within Acute Myeloid Leukemia. Another contribution is the one by \cite{salama2022artificial}, where in their paper they focus on the application of artificial intelligence in enhancing the diagnostic flow cytometry workflow for the detection of minimal residual disease in Chronic Leukemia. The authors show how AI can significantly improve the accuracy and efficiency of Chronic Leukemia analysis. Additionally, the study by \cite{reiter2019automated} provides valuable insights into B Cell Acute Lymphoblastic Leukemia (B-ALL) analysis. The authors propose an automated flow cytometric MRD assessment method using supervised machine learning. Their method has shown promising results in the detection and analysis of B-ALL.

While these works have significantly advanced the field, our study takes a novel approach by applying GATs for hematological analysis. To the best of our knowledge, we are the first to use GATs for this purpose. Our approach has yielded impressive results in terms of various metrics, including micro accuracy, micro precision, micro recall, and micro f1 score. Furthermore, our method has demonstrated its effectiveness in terms of corrected ratios of labels predicted, as we will detail in the following sections. This pioneering application of GATs in multi-class flow cytometry data analysis opens up new possibilities for the use of advanced machine learning techniques in biomedical research.

\section{Architecture of Our Model}
\label{Architecture}
Our approach is based on the GATs as described by \cite{GATVel}, and we adhere to their exposition in this section. The fundamental unit of our network is the graph attention layer. This layer takes a collection of node features,
\begin{align}
    \{\mathbf{f}_1, \cdots, \mathbf{f}_i, \cdots, \mathbf{f}_n\},
\end{align} where $n$ is the total number of nodes, $\mathbf{f}_i \in \mathbb{R}^m$, with $m$ being the total number of features per node, i.e., the dimension. The layer outputs a new set of node features (which may have a different dimension $h$), 
\begin{align}
    \{\mathbf{f}^{'}_1, \cdots, \mathbf{f}^{'}_i, \cdots, \mathbf{f}^{'}_n\},
\end{align} where $\mathbf{f'}_i \in \mathbb{R}^{h}$.

We apply a linear transformation, termed the weight matrix, $\mathcal{V} \in \mathbb{R}^{h \times m}$ to each node. This matrix is initialized using Glorot initialization \cite{Glorot}. We then compute self-attention on the nodes, which is a shared attention mechanism
\begin{equation}
\label{att1}
   b: \mathbb{R}^{h}\times \mathbb{R}^{h} \to \mathbb{R}
\end{equation} that calculates attention coefficients
\begin{equation}
\label{att2coeff}
    g_{ij} = b\biggl(\mathcal{V}\mathbf{f}_i, \mathcal{V}\mathbf{f}_j \biggr) .
\end{equation}
These $g_{ij}$'s are treated as importance scores of node $j$'s features to node $i$. This is the key distinction between a GAT and a Graph Convolution Network (GCN) \cite{GCNKipf}. Unlike graph convolution, different importance scores are assigned to different nodes in the same neighborhood, allowing the model to better adapt to complex datasets like ours. 

We normalize these $g_{ij}$'s across all choices of $j$, where $j$ is in the same neighborhood as $i$ using the softmax function:
\begin{equation}
\label{softmax}
    \beta_{ij} = \text{softmax}_j(g_{ij}) = \frac{\exp{(g_{ij})}}{\sum_{k \in \mathcal{N}_i}\exp{(g_{ik})}}
\end{equation} where $\mathcal{N}_i$ is the corresponding neighborhood of node $i$. In all our applications, the graphs are k-nearest neighbor graphs and $\mathcal{N}_i$ is the first order $k$-neighbors of the node $i$. Thus in practice, we only have to compute $g_{ij}$ (and thus $\beta_{ij}$) for all nodes $j$ in the first-order neighborhood of node $i$, instead of all pairs of nodes $i$ and $j$. 

In our experiments, the attention mechanism $b$ is a single-layer feedforward neural network, and we apply the LeakyReLU activation function as non-linear terms for the attention. Then, the normalized attention coefficients $\beta$ are used to compute a linear combination of the features corresponding to them, to serve as the final output features vector for every node (after applying a nonlinearity):
\begin{equation}
\label{gat_layer}
    \mathbf{f'}_i = \text{ReLU}\biggl(\sum_{j \in \mathcal{N}_i}\beta_{ij}\mathcal{V}\mathbf{f}_{j}\biggr) .
\end{equation}
Following the advances in attention-based networks, we used multi-head attention similar to \cite{vaswani2017attention} and \cite{GATVel}. In this case, each attention head learns a unique set of attention weights independent of the other heads in a given layer. Specifically, $K$ independent attention mechanisms execute the transformation of Eq. \ref{gat_layer}, and then their features are concatenated, resulting in the following output feature representation:
\begin{equation}
\label{multihead}
    \mathbf{f'}_i = \biggl|\biggl|_{l=1}^{K}\text{ReLU}\biggl(\sum_{j \in \mathcal{N}_i}\beta_{ij}^{l}\mathcal{V}^{l}\mathbf{f}_j \biggr) ,
\end{equation} where we denote with $||$ the vector concatenation operation, with $\beta_{ij}^{l}$ the normalized attention coefficients computed by the $l$-th attention mechanism and with $\mathcal{V}^{l}$ the corresponding input linear transformation's weight matrix. In this way, in the final output layer, each node will have exactly $Kh$ features.

Finally, in the terminal layer (the prediction layer) of the graph network, we first apply an averaging operation and then apply the final nonlinearity (here logarithmic softmax function, given that our task is a classification problem). Hence, the equation for the final layer is:
\begin{equation}
\label{final_layer}
    \hat{y}_i = \text{log\textunderscore softmax}\biggl(\frac{1}{K}\sum_{l=1}^{K}\sum_{j \in \mathcal{N}_i}\beta_{ij}^l\mathcal{V}^l\mathbf{f}_j \biggr) .  
\end{equation}  To prevent the gradients from exploding, we have also employed weight clipping. Table \ref{table_hype} presents all the hyperparameters in our model that were used for the experiments. Moreover, to address the challenge posed by the strong population imbalance within our dataset, we integrated a negative log-likelihood weighted loss function into our model, computing different weights for every class of every patient. 
One of the reasons we chose to work with GATs is its interpretability power. It is straightforward to visualize the attention heads, i.e., the attention scores between various nodes by a head in any given layer.

\section{Experiments}
\label{Experiments}
In the following sections, we delve into the specifics of our experimental setup, which incorporates both inductive and transductive learning tasks. This dual approach allows us to leverage the strengths of both learning paradigms, providing a comprehensive analysis of our data, and enhancing the robustness and reliability of our results. 

Our process begins with each patient’s data, which is represented as a tabular matrix (see Section 2.1) with approximately hundreds of thousands of rows and twelve features, as described in the supplementary material. Each row in this matrix corresponds to a cell (it will be a node in the graph space), and the twelve features represent various measurements for each cell, described in Table \ref{tab:1}.

We then identify $\mathcal{N}_i$ as the set of $k$-nearest neighbor for node $i$ (here $k=7$) using $\text{L}_2$ metric in the feature space. For each node, we build all combinations of pairs of its neighbors and add them to the edge set of the graph. This process results in a fully connected graph for each instance. After different trials, we chose to use $k=7$ to strike a balance between making the graph too sparse (which could miss important connections) and too dense (which could include irrelevant connections). This process allows us to capture both the local structure (through the individual neighbors) and the higher-order relationships (through the combinations of neighbors), helping us capture the most relevant relationships in the data without adding too much noise. Moreover, constructing the graph via the $k$-nearest neighbor is intrinsically local and does not force us to make any assumptions about the distribution of the data, making it a good choice for our dataset where the full distribution of the data is unknown.

\subsection{Inductive Learning}
\label{Inductive}
For this specific task, we utilized the dataset from our cohort of 30 patients. To ensure a robust evaluation methodology, we adopted a randomized approach. Specifically, to evaluate the predicted labels $\hat{y}_i$ by our model, we performed a 7-fold test procedure and 10-fold cross validation within training, meaning out of 30 patients, 4 or 5 are for testing, 2 or 3 for validation and the rest for training. It is important to note that the testing graphs are completely unseen during training.

We train our model and tune the hyperparameters on the validation set to evaluate the performance on the test set. We ran our HemaGraph model 7 times with 7 different seed initializations to ensure the robustness of our results.

The performance is measured in terms of accuracy, precision, recall, and F1 score as reported in Table \ref{tab:2}, where we compared HemaGraph with state-of-the-art methods for tabular data classification, like Deep Neural Network (DNN), XGBoost (XGB), Random Forest (RF), Gaussian Mixture Models (GMM), and with popular GNNs like GraphSAGE (SAGE) \cite{GraphSAGE} and GCN \cite{GCNKipf}.

We computed these metrics for each class $i$, using the following formula (where TP stands for True Positive, FN for False Negative, and FP for False Positive):

\begin{equation*}
\begin{split}
\text{Precision}_i &= \frac{\text{TP}_i}{\text{TP}_i + \text{FP}_i}, \\
\text{Recall}_i &= \frac{\text{TP}_i}{\text{TP}_i + \text{FN}_i}, \\
\text{F1 Score}_i &= 2 \times \frac{\text{Precision}_i \times \text{Recall}_i}{\text{Precision}_i + \text{Recall}_i}.
\end{split}
\end{equation*} We then would have similar equations for $i = 1, 2, 3, 4, 5$. Encouraged by the predictive ability of our model, we take a deeper look, and in Table \ref{tab:3a} and Table \ref{tab:3b}  we show the (correct) predicted label across patients and cell types, compared to the state-of-the-art multi-class classification methods for tabular data.

\begin{table}
    \footnotesize
    \centering
    \begin{tabular}{ccccc}
        \hline
        Model & Accuracy & Precision & Recall & F1-Score \\
        \hline
        \textbf{HemaGraph} & \textbf{0.98} & \textbf{0.99} & \textbf{0.98} & \textbf{0.98} \\
        SAGE & 0.93 & 0.93 & 0.92 & 0.91 \\
        GCN & 0.90 & 0.92 & 0.90 & 0.89 \\
        DNN & 0.86 & 0.88 & 0.84 & 0.80 \\
        RF & 0.84 & 0.81 & 0.83 & 0.83 \\
        XGB & 0.86 & 0.85 & 0.84 & 0.85 \\
        GMM & 0.58 & 0.57 & 0.57 & 0.57 \\
        \hline
    \end{tabular}
    \caption{Average metrics across all patients, compared to the SOTA multi-class classification methods. For the sake of space, all the results are averaged with $\pm 0.01$, and the names are shortened.}
    \label{tab:2}
\end{table} 

\begin{figure}[h]
\centering \includegraphics[width=0.65\linewidth]{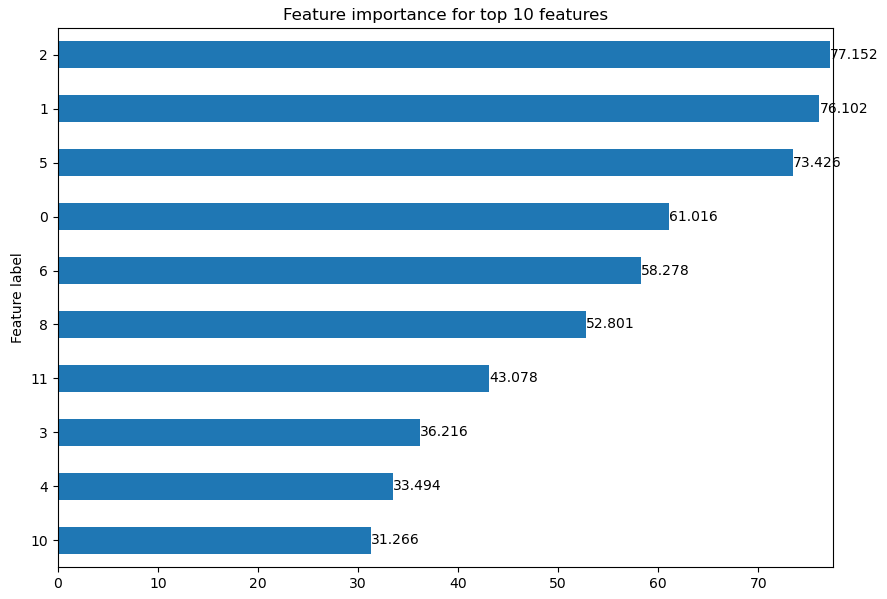}
\caption{Feature importance as highlighted by our model. For the sake of simplicity, we remind the reader to match the labels with the corresponding Table \ref{tab:1} for future explanations.}
\label{feature_importance}
\end{figure}

Moreover, we also compared our model with other graph deep learning classifiers and showed the benefit of using attention to correctly capture low-populated cell types, as in Table \ref{tab:3b} where we kept the best hyperparameter for every model for a fair comparison. In Appendix \ref{appendix2}, Table \ref{table_hype} shows all the hyperparameters that have been used throughout the experiments. Using GATs also allows us to interpret the choices made by the model, analyzing which features were considered more important than others, as shown in Figure \ref{feature_importance} where the top 10 features are shown.
 As we can see, the features that are given the most importance (at least above 55\%) are 2,1,5,0 and 6, therefore we broke down the choice of the model for each one. In order of importance, we had:
\begin{table}[H]
\footnotesize
    \centering
    \begin{tabular}{cccccc}
        \hline
        Type & \textbf{HemaGraph} & DNN & RF & XGB & GMM\\
        \hline
        T Cells & \textbf{98.81} & 96.71 & 94.60 & 97.01 & 92.97 \\
        B Cells  & \textbf{95.94} & 73.62 & 88.16 & 92.63 & - \\
        Monocytes  & \textbf{98.01} & 90.28 & 85.38 & 81.23 & 88.78\\
        Mast  & \textbf{95.43} & - & - & 0.44 & - \\
        HSPC & \textbf{83.77} &  5.1 & 0.06 & 4.76 & -\\
        \hline
    \end{tabular}
    \caption{Comparison with deep tabular models of average correct ratios across all 30 patients (for the sake of space, all the results are averaged with $\pm 0.01$ and the names shortened).}
    \label{tab:3a}
\end{table}
\begin{figure*}
\begin{subfigure}{.5\textwidth}
  \centering  \includegraphics[width=.9\linewidth]{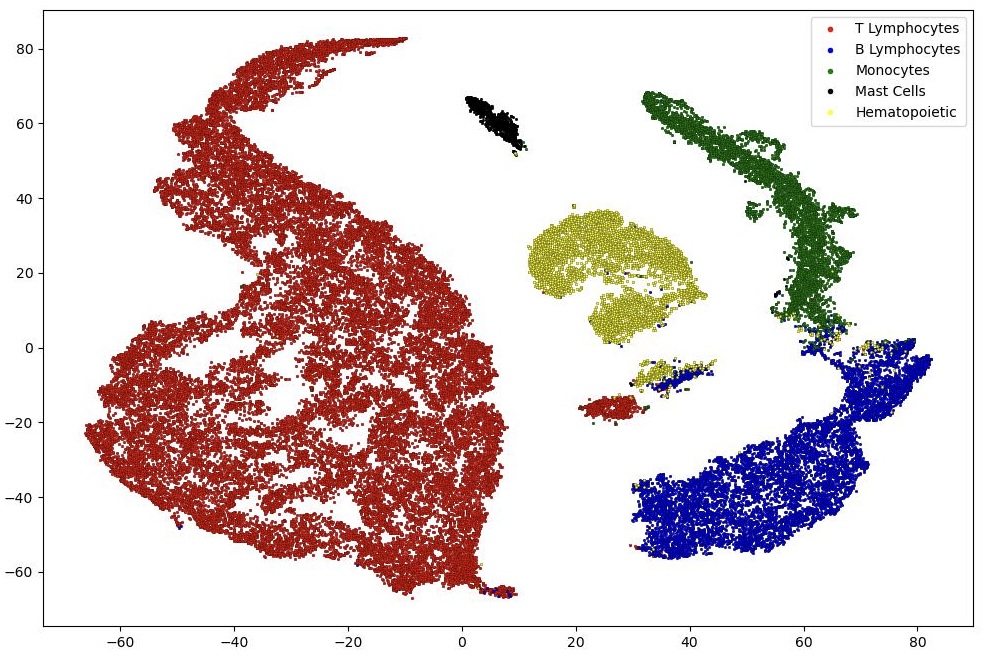} 
  \caption{t-SNE projection for patient 23, for the inductive learning task.}
 \label{ALLIN1}
\end{subfigure}
\begin{subfigure}{.5\textwidth}
 \centering
 \includegraphics[width=0.65\linewidth]{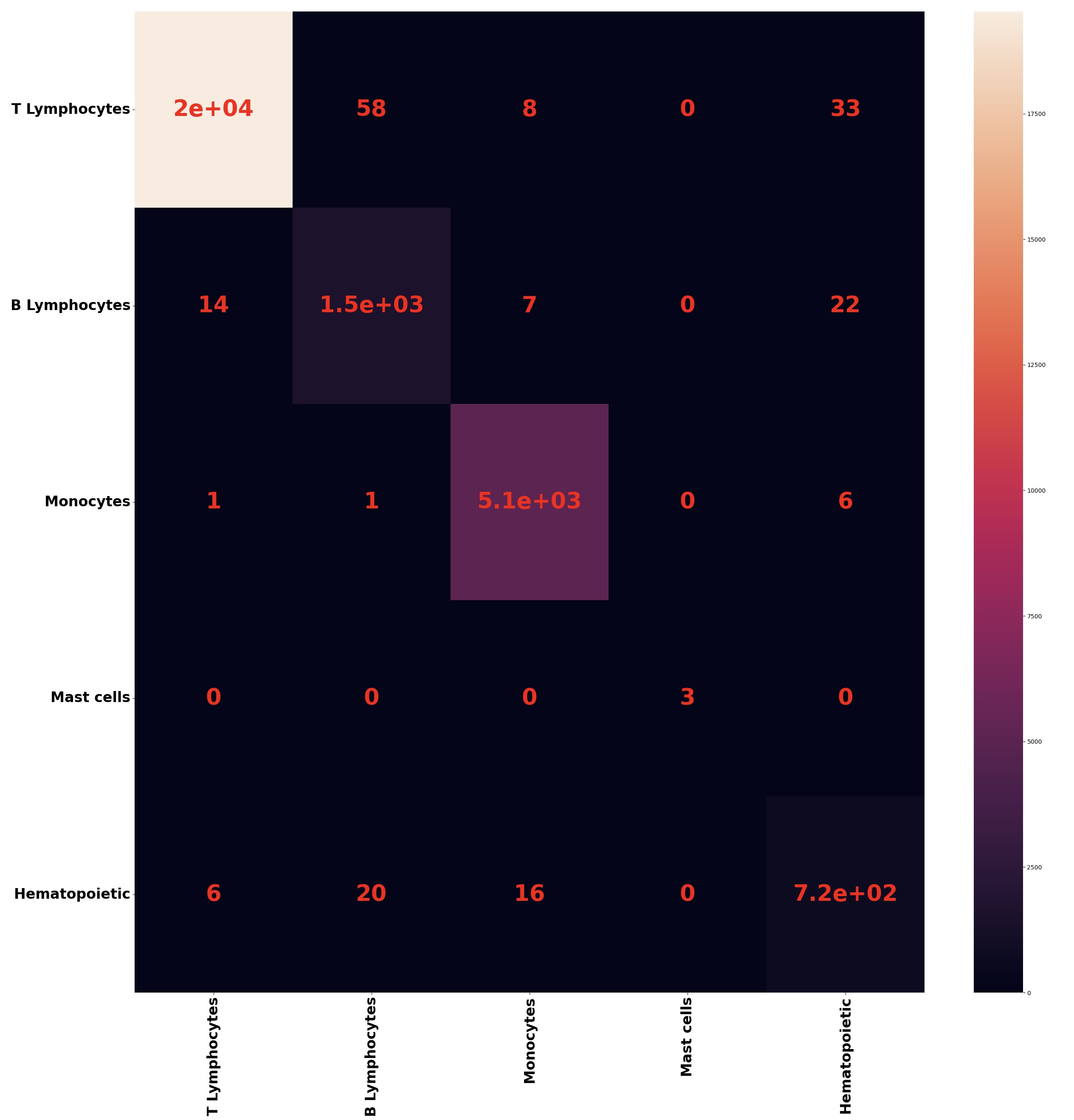} 
  \caption{Confusion matrix for patient 23.}
  \label{ALLIN2}
\end{subfigure}
\caption{Results for patient 23.}
\label{fig:fig11}
\end{figure*}
\begin{table}[H]
\footnotesize
    \centering
    \begin{tabular}{cccccc}
        \hline
        Type & \textbf{HemaGraph (ours)} & SAGE & GCN \\
        \hline
        T Cells  & \textbf{98.81} & 98.53 & 97.46\\
        B Cells  & \textbf{95.54} & 92.03 & 89.62\\
        Monocytes  & \textbf{97.91} & 94.82 & 95.92\\
        Mast Cells  & \textbf{95.00} & 79.04 & 78.18\\
        HSPC  & \textbf{83.77} & 76.84 & 76.94\\
        \hline
    \end{tabular}
    \caption{Comparison with graph deep learning classifiers of average correct ratios across all 30 patients (for the sake of space, all the results are averaged with $\pm 0.01$).}
    \label{tab:3b}
\end{table}
\begin{itemize}
    \item CD14-FITC. This feature indicates the expression level of CD14, a cell surface receptor that binds to lipopolysaccharide (LPS), a component of the bacterial cell wall. CD14 is mainly expressed by monocytes, macrophages, and activated granulocytes, and it mediates the immune response to bacterial infection. A high importance of this feature may suggest that our model can distinguish between cells that are involved in innate immunity and those that are not, detecting the presence of bacterial infection in the sample.
    \item Side Scatter (SSC)-Cell’s granularity. This feature indicates the level of light scatter at a 90-degree angle relative to the laser beam. SSC reflects the internal complexity or granularity of the cell, such as the presence of granules, nuclei, or other organelles. A high importance of this feature may suggest that our model can differentiate between cells that have different degrees of complexity, such as lymphocytes, monocytes, and granulocytes, or that your model can identify cells that have abnormal granularity, such as blast cells or malignant cells.
    \item CD33-PC5.5. This feature indicates the expression level of CD33, a cell surface receptor that belongs to the sialic acid-binding immunoglobulin-like lectin (Siglec) family. CD33 is expressed by myeloid cells, such as monocytes, macrophages, granulocytes, and mast cells, and it modulates the immune response by inhibiting the activation of these cells. A high importance of this feature may suggest that our model can distinguish between myeloid and non-myeloid cells, or that your model can detect the expression of CD33 as a marker for certain types of leukemia, such as Acute Myeloid Leukemia (AML) or chronic myelomonocytic leukemia (CMML).
    \item Forward scatter (FSC)-Cell’s size. This feature indicates the level of light scatter along the path of the laser beam. FSC is proportional to the diameter or surface area of the cell, and it can be used to discriminate cells by size. Having importance on this feature may suggest that our model can differentiate between cells that have different sizes, such as small lymphocytes and large monocytes.
    \item CD34-PC7. This feature indicates the expression level of CD34, a cell surface glycoprotein that belongs to the sialomucin family. CD34 is expressed by hematopoietic stem cells and progenitor cells (HSPCs). CD34 functions as a cell-cell adhesion molecule. Putting importance on this feature suggests that the model can identify the presence of HSPSc in the sample.
\end{itemize}
Once again, inspired by the classification power of HemaGraph, we took a deeper look at the model's performance, and we computed the confusion matrix for every patient to highlight the strong power of the model in capturing low-concentrated type cells, as shown in Figure \ref{ALLIN2} where the results of one typical patient are shown (patient 23). Moreover, to allow better visualization of the predicted cell patterns, for the same patient we show in Figure \ref{ALLIN1} the obtained t-SNE embeddings.
\begin{figure}[h]
\centering \includegraphics[width=1.0\linewidth]{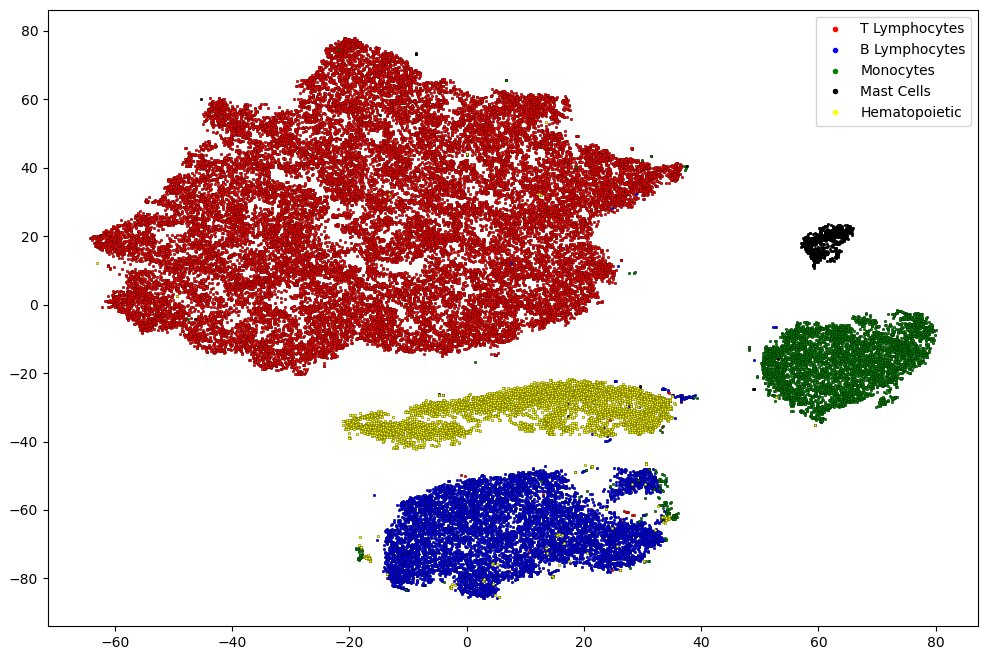}
\caption{t-SNE projection for patient 23, for the transductive learning task.}
\label{umap}
\end{figure}

\subsection{Transductive Learning}
\label{Transductive}
In our experiments, we employed once again our single-cell dataset and adhered to the transductive learning methodology as outlined in \cite{yang2016}. Within this framework, it is pertinent to note that during the training phase, our model was provided access to the feature vectors of all nodes. The primary objective of our investigation was to ascertain the status of cells, specifically differentiating between our five different classes, and to verify the ability of our attention model to get out from hundreds of thousands of nodes for some of the patients. 

To achieve this, we conducted a series of 7 different seed iterations utilizing our HemaGraph model, ensuring thorough evaluation and consistency in outcomes. Here, we randomly assign 10\% of the nodes for validation and 10\% for testing while keeping the classes' ratio balanced, the same as in the full dataset. Since our model is strongly imbalanced, we deployed a weighted negative log-likelihood loss function here as well. Then we randomly mask the labels of 50\% of the training, validation, and test set. 

Thus we can think of our problem as having a large graph where half of the graph is unlabeled and our goal is to predict the labels based on the ones that we have. Table \ref{tab4} shows the results in terms of our metrics evaluation for HemaGraph, compared with the best models (inherited from the Inductive Task) where we choose the hyperparameters (Table \ref{hyper_trans} in Appendix \ref{appendix2}) to fit the constraints of 1 GPU with 11GB memory. Moreover, as we did in Section \ref{Inductive}, we dive deeper into our model details, and Figure \ref{umap} shows the t-SNE \cite{t-SNE} embeddings visualization of HemaGraph for patient 23. We can see clearly how the model has clustered the five classes into its final embedding layer despite the difficulties due to the mask setup, typical of transductive learning.
\begin{table}
    \footnotesize
    \centering
    \begin{tabular}{ccccc}
        \hline
        Model & Accuracy & Precision & Recall & F1-Score \\
        \hline
        \textbf{HemaGraph} & \textbf{0.98} & \textbf{0.99} & \textbf{0.98} & \textbf{0.98} \\
        SAGE & 0.95 & 0.94 & 0.94 & 0.94 \\
        GCN & 0.96 & 0.95 & 0.96 & 0.95 \\
        \hline
    \end{tabular}
    \caption{Average metrics across all patients for transductive learning, compared to the state-of-the-art graph classifier. For the sake of space, all the results are averaged with $\pm 0.01$.}
    \label{tab4}
\end{table} 


\section{Conclusion and Future Directions}
\label{Conclusion}
In this paper, we introduced HemaGraph, a newly developed Graph Attention model tailored for multi-class classification problems in both inductive and transductive learning scenarios. Our comprehensive evaluation revealed that HemaGraph consistently outperforms existing state-of-the-art methods, particularly on FCS tabular data, setting a new standard in this domain.

A pivotal feature of HemaGraph is its adeptness in addressing the inherent imbalances observed across diverse cell populations. By incorporating a designed weighted loss structure, we have ensured equitable representation within the model. This strategic approach not only enhances performance but also fortifies the model against the intricate challenges posed by imbalances and complexities in the data landscape.

Looking ahead, we plan to enlarge the dataset from our patients to incorporate diverse hematologic diseases as well as other cellular populations, for example emphasizing AML ill-patient data acquisition. In this way, we aim to establish a robust benchmark \cite{flowcyt2024} that encapsulates the intricacies of cellular analysis comprehensively. Moreover, we will focus our work on unlocking the potential of semi-supervised and self-supervised learning methodologies. These approaches are inherently less time-consuming (fewer labels needed) and financially intensive, offering a pragmatic avenue for scalable advancements.

Furthermore, it's imperative to underscore the potential of AI (HemaGraph particularly) in clinical settings. The speed and efficiency of our model are a huge advantage. While it typically takes 20 to 25 minutes to manually analyze flow cytometry data obtained from a patient, HemaGraph accomplishes the same task in a mere 1-minute inference time. 

In conclusion, the integration of AI, exemplified by HemaGraph, into routine medical practices heralds a paradigm shift. As we continue to refine and expand our methodologies, the main goal remains clear: to develop AI models that are not only robust and reliable but also well integrated into the clinical workflow, empowering physicians and enhancing patient care.

\appendix
\section{Hyperparameters for Tabular Data Classifiers}
\label{appendix2}
Table \ref{table_hype} and Table \ref{hyper_trans} below show the default hyperparameters used in all the experiments:
\begin{table}[H]
    \centering
    \resizebox{0.40\textwidth}{!}{%
    \begin{tabular}{ccccc}
        \hline
        Parameter & HemaGraph & SAGE & GCN & DNN \\
        \hline
        Layers  &3 & 3 & 3 &3 \\
        Hidden-ch  &64 &64 & 64 &256\\
        Att-heads &8 &- & -  &-\\
        Optimizer &Adam &Adam & Adam  &Adam\\
        Lr-sched  & 0.01-1e-7 &0.01-1e-7 &0.01-1e-7 &0.01-1e-7\\
        Weigh-decay &0.0005 &0.005 & 0.005 &0.005\\
        Dropout  &0.3 &0.4 & 0.4  &0.5\\
        Epochs  &1000 &1000 & 1000  &1000\\
        Early-stop  &40 &50 & 50  &40\\
        \hline     
\end{tabular}
}
    \caption{Default hyperparameters used in all inductive experiments. For XGBoost, Random Forest, and Gaussian Mixture Model we used the online available sklearn package.} 
    \label{table_hype}
\end{table}
\begin{table}[H]
    \centering
    \resizebox{0.38\textwidth}{!}{%
    \begin{tabular}{cccc}
        \hline
        Parameter & HemaGraph & SAGE & GCN\\
        \hline
        Layers  &3 & 3 & 3\\
        Hidden-ch  &8 & 16 & 16\\
        Att-heads &8 &- & -\\
        Optimizer &Adam &Adam & Adam\\
        Lr scheduler  & 0.01-1e-7 &0.01-1e-7\\
        Weight decay &0.0005 &0.005 & 0.005\\
        Dropout  &0.3 &0.5 & 0.5\\
        Train epochs  &1000 &1000 & 1000\\
        Early stopping  &40 &50 & 50\\
        \hline
\end{tabular}
}
    \caption{Default hyperparameters used in all transductive experiments to fit our memory GPU constraints.}
    \label{hyper_trans}
\end{table}

\section*{Acknowledgments}
The Swiss National Science Foundation partially funds this work under grant number 207509 "Structural Intrinsic Dimensionality".


\bibliographystyle{named}
\bibliography{ijcai24}

\begin{thebibliography}{}

\bibitem[\protect\citeauthoryear{Bini \bgroup \em et al.\egroup }{2024}]{flowcyt2024}
Lorenzo Bini, Fatemeh Nassajian~Mojarrad, Margarita Liarou, Thomas Matthes, and Stephane Marchand-Maillet.
\newblock Flowcyt: A comparative study of deep learning approaches for multi-class classification in flow cytometry.
\newblock 2024.

\bibitem[\protect\citeauthoryear{Bove \bgroup \em et al.\egroup }{2020}]{bove2020prediction}
Pasquale Bove, Alessio Micheli, Paolo Milazzo, Marco Podda, et~al.
\newblock Prediction of dynamical properties of biochemical pathways with graph neural networks.
\newblock In {\em Bioinformatics}, pages 32--43, 2020.

\bibitem[\protect\citeauthoryear{Cai \bgroup \em et al.\egroup }{2017}]{cai2017}
Han Cai, Kan Ren, Weinan Zhang, Kleanthis Malialis, Jun Wang, Yong Yu, and Defeng Guo.
\newblock Real-time bidding by reinforcement learning in display advertising.
\newblock In {\em Proceedings of the tenth ACM international conference on web search and data mining}, pages 661--670, 2017.

\bibitem[\protect\citeauthoryear{Gao \bgroup \em et al.\egroup }{2022}]{gao2022graph}
Chen Gao, Xiang Wang, Xiangnan He, and Yong Li.
\newblock Graph neural networks for recommender system.
\newblock In {\em Proceedings of the Fifteenth ACM International Conference on Web Search and Data Mining}, pages 1623--1625, 2022.

\bibitem[\protect\citeauthoryear{Glorot and Bengio}{2010}]{Glorot}
Xavier Glorot and Yoshua Bengio.
\newblock Understanding the difficulty of training deep feedforward neural networks.
\newblock In {\em Proceedings of the thirteenth international conference on artificial intelligence and statistics}, pages 249--256. JMLR Workshop and Conference Proceedings, 2010.

\bibitem[\protect\citeauthoryear{Grinsztajn \bgroup \em et al.\egroup }{2022}]{grinsztajn2022tree}
L{\'e}o Grinsztajn, Edouard Oyallon, and Ga{\"e}l Varoquaux.
\newblock Why do tree-based models still outperform deep learning on typical tabular data?
\newblock {\em Advances in Neural Information Processing Systems}, 35:507--520, 2022.

\bibitem[\protect\citeauthoryear{Hamilton \bgroup \em et al.\egroup }{2017}]{GraphSAGE}
Will Hamilton, Zhitao Ying, and Jure Leskovec.
\newblock Inductive representation learning on large graphs.
\newblock {\em Advances in neural information processing systems}, 30, 2017.

\bibitem[\protect\citeauthoryear{He \bgroup \em et al.\egroup }{2017}]{he2017neural}
Xiangnan He, Lizi Liao, Hanwang Zhang, Liqiang Nie, Xia Hu, and Tat-Seng Chua.
\newblock Neural collaborative filtering.
\newblock In {\em Proceedings of the 26th international conference on world wide web}, pages 173--182, 2017.

\bibitem[\protect\citeauthoryear{Kipf and Welling}{2016}]{GCNKipf}
Thomas~N Kipf and Max Welling.
\newblock Semi-supervised classification with graph convolutional networks.
\newblock {\em arXiv preprint arXiv:1609.02907}, 2016.

\bibitem[\protect\citeauthoryear{Li \bgroup \em et al.\egroup }{2019}]{li2019fi}
Zekun Li, Zeyu Cui, Shu Wu, Xiaoyu Zhang, and Liang Wang.
\newblock Fi-gnn: Modeling feature interactions via graph neural networks for ctr prediction.
\newblock In {\em Proceedings of the 28th ACM international conference on information and knowledge management}, pages 539--548, 2019.

\bibitem[\protect\citeauthoryear{Reiter \bgroup \em et al.\egroup }{2019}]{reiter2019automated}
Michael Reiter, Markus Diem, Angela Schumich, Margarita Maurer-Granofszky, Leonid Karawajew, Jorge~G Rossi, Richard Ratei, Stefanie Groeneveld-Krentz, Elisa~O Sajaroff, Susanne Suhendra, et~al.
\newblock Automated flow cytometric mrd assessment in childhood acute b-lymphoblastic leukemia using supervised machine learning.
\newblock {\em Cytometry Part A}, 95(9):966--975, 2019.

\bibitem[\protect\citeauthoryear{Rendle}{2010}]{rendle2010factorization}
Steffen Rendle.
\newblock Factorization machines.
\newblock In {\em 2010 IEEE International conference on data mining}, pages 995--1000. IEEE, 2010.

\bibitem[\protect\citeauthoryear{Salama \bgroup \em et al.\egroup }{2022}]{salama2022artificial}
Mohamed~E Salama, Gregory~E Otteson, Jon~J Camp, Jansen~N Seheult, Dragan Jevremovic, David~R Holmes~III, Horatiu Olteanu, and Min Shi.
\newblock Artificial intelligence enhances diagnostic flow cytometry workflow in the detection of minimal residual disease of chronic lymphocytic leukemia.
\newblock {\em Cancers}, 14(10):2537, 2022.

\bibitem[\protect\citeauthoryear{Shan \bgroup \em et al.\egroup }{2023}]{shan2023glae}
Yixiang Shan, Jielong Yang, Xiangtao Li, Xionghu Zhong, and Yi~Chang.
\newblock Glae: A graph-learnable auto-encoder for single-cell rna-seq analysis.
\newblock {\em Information Sciences}, 621:88--103, 2023.

\bibitem[\protect\citeauthoryear{Song \bgroup \em et al.\egroup }{2019}]{song2019autoint}
Weiping Song, Chence Shi, Zhiping Xiao, Zhijian Duan, Yewen Xu, Ming Zhang, and Jian Tang.
\newblock Autoint: Automatic feature interaction learning via self-attentive neural networks.
\newblock In {\em Proceedings of the 28th ACM international conference on information and knowledge management}, pages 1161--1170, 2019.

\bibitem[\protect\citeauthoryear{Thais \bgroup \em et al.\egroup }{2022}]{thais2022graph}
Savannah Thais, Paolo Calafiura, Grigorios Chachamis, Gage DeZoort, Javier Duarte, Sanmay Ganguly, Michael Kagan, Daniel Murnane, Mark~S Neubauer, and Kazuhiro Terao.
\newblock Graph neural networks in particle physics: Implementations, innovations, and challenges.
\newblock {\em arXiv preprint arXiv:2203.12852}, 2022.

\bibitem[\protect\citeauthoryear{Van~der Maaten and Hinton}{2008}]{t-SNE}
Laurens Van~der Maaten and Geoffrey Hinton.
\newblock Visualizing data using t-sne.
\newblock {\em Journal of machine learning research}, 9(11), 2008.

\bibitem[\protect\citeauthoryear{Vaswani \bgroup \em et al.\egroup }{2017}]{vaswani2017attention}
Ashish Vaswani, Noam Shazeer, Niki Parmar, Jakob Uszkoreit, Llion Jones, Aidan~N Gomez, {\L}ukasz Kaiser, and Illia Polosukhin.
\newblock Attention is all you need.
\newblock {\em Advances in neural information processing systems}, 30, 2017.

\bibitem[\protect\citeauthoryear{Velickovic \bgroup \em et al.\egroup }{2017}]{GATVel}
Petar Velickovic, Guillem Cucurull, Arantxa Casanova, Adriana Romero, Pietro Lio, Yoshua Bengio, et~al.
\newblock Graph attention networks.
\newblock {\em stat}, 1050(20):10--48550, 2017.

\bibitem[\protect\citeauthoryear{Wang \bgroup \em et al.\egroup }{2020}]{wang2020graph}
Shoujin Wang, Liang Hu, Yan Wang, Xiangnan He, Quan~Z Sheng, Mehmet Orgun, Longbing Cao, Nan Wang, Francesco Ricci, and Philip~S Yu.
\newblock Graph learning approaches to recommender systems: A review.
\newblock {\em arXiv preprint arXiv:2004.11718}, 2020.

\bibitem[\protect\citeauthoryear{Wang \bgroup \em et al.\egroup }{2021a}]{scgnn}
Juexin Wang, Anjun Ma, Yuzhou Chang, Jianting Gong, Yuexu Jiang, Ren Qi, Cankun Wang, Hongjun Fu, Qin Ma, and Dong Xu.
\newblock scgnn is a novel graph neural network framework for single-cell rna-seq analyses.
\newblock {\em Nature communications}, 12(1):1882, 2021.

\bibitem[\protect\citeauthoryear{Wang \bgroup \em et al.\egroup }{2021b}]{wang2021dcn}
Ruoxi Wang, Rakesh Shivanna, Derek Cheng, Sagar Jain, Dong Lin, Lichan Hong, and Ed~Chi.
\newblock Dcn v2: Improved deep \& cross network and practical lessons for web-scale learning to rank systems.
\newblock In {\em Proceedings of the web conference 2021}, pages 1785--1797, 2021.

\bibitem[\protect\citeauthoryear{Weijler \bgroup \em et al.\egroup }{2022}]{weijler2022umap}
Lisa Weijler, Florian Kowarsch, Matthias W{\"o}dlinger, Michael Reiter, Margarita Maurer-Granofszky, Angela Schumich, and Michael~N Dworzak.
\newblock Umap based anomaly detection for minimal residual disease quantification within acute myeloid leukemia.
\newblock {\em Cancers}, 14(4):898, 2022.

\bibitem[\protect\citeauthoryear{Wu \bgroup \em et al.\egroup }{2020}]{wu2020comprehensive}
Zonghan Wu, Shirui Pan, Fengwen Chen, Guodong Long, Chengqi Zhang, and S~Yu Philip.
\newblock A comprehensive survey on graph neural networks.
\newblock {\em IEEE transactions on neural networks and learning systems}, 32(1):4--24, 2020.

\bibitem[\protect\citeauthoryear{Yang \bgroup \em et al.\egroup }{2016}]{yang2016}
Zhilin Yang, William Cohen, and Ruslan Salakhudinov.
\newblock Revisiting semi-supervised learning with graph embeddings.
\newblock In {\em International conference on machine learning}, pages 40--48. PMLR, 2016.

\bibitem[\protect\citeauthoryear{Yun \bgroup \em et al.\egroup }{2021}]{yun2021neo}
Seongjun Yun, Seoyoon Kim, Junhyun Lee, Jaewoo Kang, and Hyunwoo~J Kim.
\newblock Neo-gnns: Neighborhood overlap-aware graph neural networks for link prediction.
\newblock {\em Advances in Neural Information Processing Systems}, 34:13683--13694, 2021.

\bibitem[\protect\citeauthoryear{Zhou \bgroup \em et al.\egroup }{2020}]{zhou2020graph}
Jie Zhou, Ganqu Cui, Shengding Hu, Zhengyan Zhang, Cheng Yang, Zhiyuan Liu, Lifeng Wang, Changcheng Li, and Maosong Sun.
\newblock Graph neural networks: A review of methods and applications.
\newblock {\em AI open}, 1:57--81, 2020.

\end{thebibliography}
\end{document}